\documentclass[%
 reprint,
 superscriptaddress,
 amsmath,amssymb,
 aps,
prab,
floatfix,
]{revtex4-2}
\usepackage[english]{babel}
\usepackage{graphicx}   
\usepackage{dcolumn}    
\usepackage{bm}         

\usepackage[table,dvipsnames]{xcolor}
\usepackage{booktabs}
\usepackage[separate-uncertainty=true]{siunitx}
\usepackage{hyperref}
\usepackage{amsmath}
\usepackage[capitalize]{cleveref}
\usepackage{float}
\usepackage{amssymb}
\usepackage{threeparttable}
\usepackage{xcolor}

\usepackage[acronym]{glossaries}
\glsdisablehyper

\newacronym{AD}{AD}{auto-differentiation}
\newacronym{APS}{APS}{Advanced Photon Source}
\newacronym{BPM}{BPM}{beam position monitor}
\newacronym{BTS}{BTS}{Booster-to-Storage-ring}
\newacronym{CNSGA}{CNSGA}{chaotic non-dominated sorting genetic algorithm}
\newacronym{GPSR}{GPSR}{generative phase space reconstruction}
\newacronym{MAE}{MAE}{mean absolute error}
\newacronym{ML}{ML}{machine learning}
\newacronym{MSE}{MSE}{mean squared error}
\newacronym{NN}{NN}{neural network}
\newacronym[longplural={transverse deflecting cavities}]{TCAV}{TCAV}{transverse deflecting cavity}
\newacronym{RMS}{RMS}{root mean square}

\usepackage{mathtools}
\usepackage{listings}  
\lstdefinestyle{myStyle}{
    aboveskip={1.3\baselineskip},
    basicstyle=\footnotesize\ttfamily\linespread{4},
    breaklines=false,
    columns=flexible,
    commentstyle=\color[rgb]{0.127,0.427,0.514}\ttfamily\itshape,
    escapechar=@,
    extendedchars=true,
    frame=single,
    identifierstyle=\color{black},
    inputencoding=latin1,
    keywordstyle=\color[HTML]{228B22}\bfseries,
    ndkeywordstyle=\color[HTML]{228B22}\bfseries,
    prebreak = \raisebox{0ex}[0ex][0ex]{\ensuremath{\hookleftarrow}},
    showstringspaces=false,
    stringstyle=\color[rgb]{0.639,0.082,0.082}\ttfamily,
    upquote=true,
}

\begin{document}

\preprint{APS/123-QED}

\title{Five-Dimensional Beam Sigma Matrix Determination in Transport Lines with Differentiable Simulation}

\author{Chenran Xu}
\email{chenran.xu@anl.gov}
\author{Louis Emery}
\author{Osama Mohsen}
\affiliation{Argonne National Laboratory, Lemont, IL, USA}
\author{Ryan Roussel}
\affiliation{SLAC National Accelerator Laboratory, Menlo Park, CA, USA}
\author{Kent P. Wootton}
\author{Yine Sun}
\author{Michael Borland}
\affiliation{Argonne National Laboratory, Lemont, IL, USA}


\date{\today}

\begin{abstract}
    Precise measurement of the beam sigma matrix is essential for matching the optics in transport lines and ensuring reliable accelerator operation. In this work, we present a method for measuring and reconstructing the non-temporal five-dimensional beam sigma matrix using quadrupole scans performed in a dispersive transport region. The proposed approach enables characterization of the beam moments using only quadrupoles and beam transverse profile diagnostics, without requiring  longitudinal diagnostics or a dedicated beamline section. To achieve robust and computationally efficient reconstruction, we formulate the problem within a differentiable simulation framework, allowing direct gradient-based optimization of the initial beam covariance matrix. We demonstrate the method experimentally in the Booster-to-Storage-ring (BTS) transport line at the Advanced Photon Source (APS), where it produces consistent reconstructions of the beam sigma matrix from measurements. We further show that the framework is flexible with respect to the number and placement of diagnostic screens, making it applicable to a broad range of existing transport-line configurations. 
    These results establish the proposed 5D beam sigma matrix reconstruction method as a practical and broadly deployable approach for fast, efficient beam characterization during accelerator operation.
\end{abstract}

\maketitle


\section{Introduction}\label{sec:introduction}

Accurate measurement of beam emittance and phase-space information is a central requirement for accelerator operation. The beam sigma matrix provides a compact description of the second-order moments of the distribution and is widely used for optics matching, transport-line model validation, injection tuning, and identifying coupling or dispersion errors. In transport lines, these measurements are particularly important because the delivered beam quality depends not only on the projected transverse emittances, but also on correlations between transverse coordinates and energy deviation.

The quadrupole-scan method is one of the simplest and most widely used approaches for emittance measurement. By varying a quadrupole strength and observing the resulting beam sizes on a downstream profile monitor, the projected emittance and Twiss parameters can be inferred from the known beam transport. With multiple quadrupoles or multiple profile monitors, the same idea can be extended to recover the full transverse four-dimensional beam sigma matrix, including $x$-$y$ cross-plane coupling terms~\cite{prat2014fourdimensional,wolski2020transverse}. This makes quadrupole scans the standard method for routine optics characterization because they can be performed with existing beamline magnets and profile diagnostics.

Longitudinal and higher-dimensional phase-space information usually requires dedicated diagnostic beamlines. A dipole spectrometer can provide energy-resolved information, a \gls{TCAV} can map longitudinal coordinates into a transverse profile, and combinations of spectrometers and \glspl{TCAV} can support time-energy or higher-dimensional phase-space measurements~\cite{emma2001transverse}. 
Recent work has demonstrated five-dimensional tomographic phase-space reconstruction by combining transverse tomography with a polarizable transverse deflecting structure~\cite{jastermerz2024fived,jastermerz2025experimental}. However, full distribution reconstruction in five or six dimensions with traditional tomography generally demands a large number of measurements, taking up to several hours of continuous operation, making it difficult to use as a routine diagnostic in existing transport lines.
\Gls{ML}-based virtual diagnostics provide a complementary route, with physics-constrained generative models recently used to predict high-dimensional phase-space densities and projected two-dimensional profiles from image and scalar measurements~\cite{scheinker2025physicsconstrained}.

In this work, we propose a simple method for measuring the five-dimensional beam sigma matrix using standard quadrupole scans in a transport-line setting, building on earlier work on the $5\times5$ sigma-matrix measurement in transport lines~\cite{borland2022promise}. The method uses standard beam profile diagnostics and existing quadrupoles in a dispersive section, avoiding the need for longitudinal diagnostic hardware or a specially configured tomography beamline. Instead of relying only on an explicitly derived first-order response matrix, we formulate the reconstruction problem using differentiable beam dynamics simulation~\cite{xu2025towards}. This enables the optimization of the incoming covariance matrix directly against measured screen moments. At the same time, it employs a physics model that can include realistic lattice elements and higher-order effects. Recent differentiable accelerator modeling tools have made this type of gradient-based reconstruction practical~\cite{kaiser2024bridging,gonzalez2023towards,qiang2023differentiable}.

Differentiable simulations have also been shown to reconstruct detailed phase-space information from a few profile measurements using the \gls{GPSR} method~\cite{roussel2023phase,roussel2024efficient,kim2024four}. In contrast, the method presented here focuses on the physically consistent reconstruction of Gaussian beam properties through the covariance matrix. This provides a simple and interpretable diagnostic for routine transport-line characterization, particularly in cases where the extracted beam is well described by its second moments, as is commonly the case for beams extracted from ring accelerators.

The paper is organized as follows. \Cref{sec:problem_formulation} introduces the sigma-matrix reconstruction problem and the differentiable-simulation method. \Cref{sec:simulation} presents simulation studies on the \gls{BTS} transport line \cite{apsu-fdr} at \gls{APS} \cite{borl:ipac18-thxgbd1}, and \cref{sec:robustness} discusses the robustness of the method when faced with realistic measurement and modeling errors.
\Cref{sec:result_bts} reports experimental reconstruction results from real-world quadrupole-scan measurements and shows that the method can be seamlessly extended to a multi-screen configuration. Finally, \cref{sec:gpsr} discusses the connection to the \gls{GPSR} method.

\section{Problem Formulation}\label{sec:problem_formulation}

The reconstruction problem is to determine the incoming covariance matrix at the entrance of a transport line from downstream screen measurements acquired while varying selected quadrupole strengths. In this work, the unknown beam state is the five-dimensional covariance matrix for the non-temporal phase space coordinates $(x,p_x,y,p_y,\delta)$. For each quadrupole setting, the measured screen image provides the transverse second moments $(\sigma_{xx},\sigma_{yy},\sigma_{xy})$ at one or more profile monitors. The inverse problem is therefore to find an incoming covariance matrix whose transported transverse moments best match these measurements over the full scan.

This problem can be approached in two related ways. The conventional method derives an explicit linear system from first-order transport matrices and solves directly for the independent entries of the incoming sigma matrix. This provides useful intuition and a natural baseline estimate, so we first summarize the linear formulation before introducing the differentiable-simulation reconstruction used in the rest of this work.

\subsection{Linear Solution}\label{sec:linear_solution}
In the linear approximation, the beam transport is treated as a linear transformation of the initial covariance matrix.
The output beam covariance matrix can be expressed as
\begin{equation}
    \Sigma_\text{out} = R \Sigma {R}^\intercal ,
\end{equation}
where $R$ is the first-order transport matrix from the reconstruction point to the downstream screen.
When the varied quadrupoles are located in a dispersive region, for example downstream of a dipole in a transport line or in a bunch-compressor chicane, they also vary the dispersion term $R_{16}$ and thereby change the measured beam sizes.
Assuming that bending occurs only in the horizontal plane, i.e.\,$R_{36}=R_{46}=0$, the transverse moments of the output beam can be expressed as
\begin{equation}
\begin{aligned}
    {(\Sigma_\text{out})}_{11} &= R_{11}^{2} {\Sigma}_{11} + 2 R_{11} R_{12} {\Sigma}_{12} + R_{12}^2 {\Sigma}_{22}  \\
    & \, + 2 R_{11} R_{16} {\Sigma}_{16} + 2 R_{12} R_{16} {\Sigma}_{26} + R_{16}^2 {\Sigma}_{66} \\
    {(\Sigma_\text{out})}_{33} &= R_{33}^{2} {\Sigma}_{33} + 2 R_{33} R_{34} {\Sigma}_{34} + R_{34}^2 {\Sigma}_{44} \\
    {(\Sigma_\text{out})}_{13} &= R_{11}R_{33} {\Sigma}_{13} + R_{11} R_{34} {\Sigma}_{14} + R_{12} R_{33} {\Sigma}_{23}  \\
    & \, + R_{12}R_{34} {\Sigma}_{24} + R_{16}R_{33} {\Sigma}_{36} + R_{16}R_{34} {\Sigma}_{46}.
\end{aligned}
\end{equation}
Stacking these equations over all quadrupole settings and screens gives an overdetermined linear system for the independent elements of $\Sigma$. A least-squares solution provides the linear estimate of the incoming beam sigma matrix.

In practical application\cite{borland2022promise}, this formulation is problematic because it neglects higher-order effects, most critically chromatic aberration from quadrupoles.
These can reduce prediction accuracy for beams exhibiting higher energy spread or in transport systems with strong focusing. This may in principle be mitigated through additional parameter-free optimization starting from the linear solution. In practice, such optimization is computationally intensive due to the large search space involved, giving slow and incomplete convergence.

\subsection{Reconstruction Using Differentiable Simulation}\label{sec:diff_reconstruction}

\begin{figure*}[htb!]
    \centering
    \includegraphics[width=0.75\linewidth]{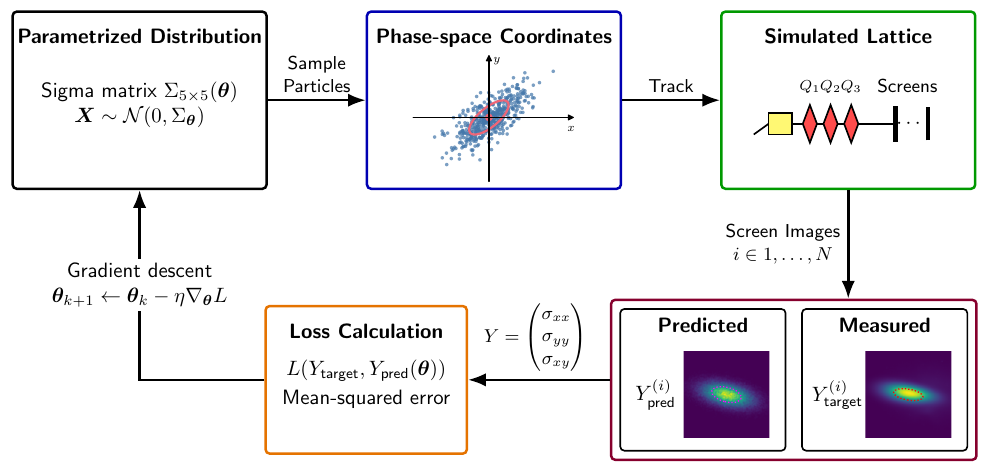}
    \caption{Workflow for the beam sigma matrix reconstruction with differentiable simulation.}
    \label{fig:workflow}
\end{figure*}

These limitations motivate the use of differentiable simulation for sigma-matrix reconstruction. Instead of relying on an explicitly derived linear response model, a differentiable simulator performs physics-based beam tracking through the accelerator lattice. This makes it possible to include higher-order effects, nonlinear elements, and other machine-specific effects as needed. At the same time, the differentiable formulation provides direct access to gradients of the reconstruction objective with respect to the initial beam moments, enabling efficient gradient-based optimization in a high-dimensional parameter space. 
The reconstruction can therefore be posed as an end-to-end optimization problem, as outlined in \cref{fig:workflow}.

The incoming beam distribution is characterized by the five-dimensional covariance matrix $\Sigma$, which can be parametrized by a lower-triangular Cholesky factor,
\begin{equation}
    \Sigma(\bm{\theta}) = L_{\bm{\theta}} L_{\bm{\theta}}^\intercal,
    \quad \bm{\theta}\in\mathbb{R}^{15},
\end{equation}
which results in 15 degrees of freedom corresponding to the independent elements of the symmetric sigma matrix.
The diagonal elements of $L_{\bm{\theta}}$ are constrained to be positive using a softplus transform,
\begin{equation}
    (L_{\bm{\theta}})_{jj} =
    \log\left(1+\exp\left(\theta_{jj,\text{raw}}\right)\right),
\end{equation}
while the off-diagonal elements are used directly as optimization variables. This parameterization ensures that the resulting covariance matrix is positive definite without requiring additional constraints in the gradient optimizer.

For each optimizer step, a fixed reference macroparticle distribution is transformed to match the parametrized covariance matrix. Let $\bm{X}_0$ be the reference distribution with centroid $\bm{\mu}_0$ and Cholesky factor $L_0$. The incoming distribution for the current trial matrix is generated as
\begin{equation}
    \bm{X}_{\text{in}}(\bm{\theta}) =
    (\bm{X}_0-\bm{\mu}_0)
    (L_{\bm{\theta}}L_0^{-1})^\intercal + \bm{\mu}_0 .
\end{equation}
Equivalently, this produces macroparticles distributed according to
\begin{equation}
    \bm{X}_{\text{in}}(\bm{\theta}) \sim
    \Sigma_{\text{in}}(\bm{\theta}).
\end{equation}
The transformed particles are tracked through the transport-line model for each quadrupole setting. For quadrupole setting $i$ and screen $s$, the predicted screen images are
\begin{equation}
    \bm{Y}^{(i,s)}(\bm{\theta}) =
    f_s\left(\bm{X}_{\text{in}}(\bm{\theta}) \mid \bm{k}_{Q,i}\right),
\end{equation}
where $f_s$ denotes tracking from the entrance of the \gls{BTS} line to screen $s$. As the reconstruction in this work assumes a nearly Gaussian beam distribution, each predicted transverse profile is reduced to its second moments,
\begin{equation}
    \bm{Y}^{(i,s)}(\bm{\theta}) =
    \left(\sigma_{xx}, \sigma_{yy}, \sigma_{xy}\right)^{(i,s)}.
\end{equation}
This scalarized objective is more computationally efficient than direct image matching while still capturing the beam sizes and transverse coupling relevant to sigma-matrix reconstruction.

The target data consist of measured moments $\bm{Y}_{\text{target}}^{(i,s)}$ for a set of quadrupole settings and diagnostic screens. The loss function is defined as the mean squared difference between the measured and predicted moments,
\begin{equation}
    l(\bm{\theta}) =
    \frac{1}{3 M N}
    \sum_{s=1}^{M}\sum_{i=1}^{N}\sum_{j=1}^{3}
    \left(
    \bm{Y}_{\text{target}, j}^{(i,s)} -
    \bm{Y}_{\text{pred}, j}^{(i,s)}(\bm{\theta})
    \right)^2 ,
\end{equation}
where $N$ is the number of quadrupole settings and $M$ is the number of screens. The multi-screen formulation adds independent constraints from different lattice locations and helps reduce degeneracy among beam sigma matrices that can produce similar moments at a single screen.

Because each operation in the workflow is differentiable, the gradient $\nabla_{\bm{\theta}}l$ is obtained directly through \gls{AD}. The Cholesky parameters can then be updated with simple gradient descent,
\begin{equation}
    \bm{\theta}_{k+1} \leftarrow
    \bm{\theta}_k -
    \eta \nabla_{\bm{\theta}}l(\bm{\theta}_k),
\end{equation}
where $\eta$ is the learning rate, or with adaptive gradient-based optimizers. In this study, we use Adam~\cite{kingma2015adam}, which adjusts the effective learning rate for each parameter and improves convergence for parameters with different numerical scales.

\section{Simulation Study}\label{sec:simulation}

\begin{figure}
    \centering
    \includegraphics[width=1\linewidth]{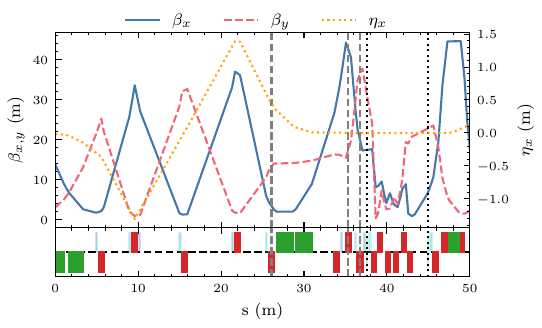}
    \caption{The \gls{BTS} transport line lattice. The three dashed lines denote the quadrupoles AQ5, BQ2, and BQ3, which are being scanned for the study. The dotted lines denote the FS3 and FS4 diagnostic screens for beam profile measurement.}
    \label{fig:bts_lattice}
\end{figure}

The proposed method is demonstrated at the upgraded \gls{BTS} transport line at \gls{APS}. The lattice of the \gls{BTS} is shown in \cref{fig:bts_lattice}. The upgraded transport line supports swap-out injection mode \cite{abela:epac92, emery:pac03-topa014} for the new storage ring at the \qty{6}{GeV} beam energy. It contains a dogleg structure that starts with two booster extraction septa, followed by a series of quadrupoles and two bending magnets. It also features an x-y emittance-exchange section consisting of six skew quadrupoles \cite{aiba:ipac15-tupje045,kuske:ipac16-weoaa01}. Proper operation of this emittance-exchange system is important for achieving high injection efficiency into the storage ring, particularly in timing modes with higher bunch charge.

Precise characterization of the booster extraction beam is therefore essential for verifying that the transport-line optics are properly matched and that the beam is delivered to the storage ring with the desired phase-space properties. To measure the five-dimensional beam sigma matrix at booster extraction, quadrupoles, before and after the bending magnets, are scanned. In this study, we use AQ5, BQ2, and BQ3, which are marked with vertical dashed lines in \cref{fig:bts_lattice}. The transverse beam profile is observed at downstream diagnostic screens. The FS3 and FS4 screens, marked as dotted lines, are of particular interest: FS3 is located closest to the varied quadrupoles, while FS4 is located directly after the x-y emittance-exchange section.

Each scan includes a set of \num{30} quadrupole settings. A large set of quadrupole strengths was randomly sampled within the allowed strength ranges, and only settings with full beam transmission were retained. 
The settings were also filtered by the predicted beam sizes $\sigma_{x,y}$ on the screens to ensure that they are not too large. Lastly, the remaining settings were down-selected to 30 with maximum variation in parameter space, as a compromise between reconstruction accuracy and required beam time for measurement. These simulations were performed with {\tt elegant} \cite{elegant,wang:pac07-thpan095}, as described in \cite{borland2022promise}.

The reconstruction method is then evaluated in simulation. The target incoming beam parameters are generated by adding \qty{5}{\percent} random perturbations to the Cholesky factors $L_\theta$ of the design covariance matrix and adding a \qty{0.1}{rad} x-y tilt.  
The obtained beam parameters are shown in \cref{tab:sim-recon-result}. The last column reports the normalized residual,
$\Delta/\sigma = (x_{\text{recon}} - x_{\text{target}})/\sigma_{\text{recon}}$,
expressed in units of the standard deviation of the repeated reconstructions. 
Overall, the reconstructed parameters are consistent with the target values, with relative errors under 1\% for most parameters except for the dispersion terms. 
The larger relative errors and normalized residuals observed for some dispersion terms are partly expected, since these parameters are small in absolute value and their contribution to the measured beam sizes is correspondingly weak.
The reconstruction uncertainty is estimated by running five optimizations with slightly varied initial guesses. The results show that the reconstruction is relatively robust to the choice of initial guess, with small variations in the final reconstructed parameters across different runs. For $\beta_x$, however, the relatively large normalized residual suggests that the statistical uncertainty estimated from repeated optimizations likely underestimates the total reconstruction uncertainty.

\begin{table}[htb!]
    \centering
    \begin{tabular}{lccc}
    \toprule
    Parameter & Target  & Reconstruction & $\Delta/\sigma$ \\\\
    \midrule
    $\epsilon_x$ (nm) & \num{69.5} & \num{70.1 \pm 0.2} & \num{3.0} \\
    $\beta_x$ (m) & \num{14.2} & \num{14.4 \pm 0.2} & \num{1.0} \\
    $\alpha_x$  & \num{2.58} & \num{2.60 \pm 0.01} & \num{1.3} \\
    $\eta_x$ (m) & \num{-0.019} & \num{-0.025 \pm 0.001} & \num{-4.2} \\
    $\eta_{x'}$ (\num{e-3}) & \num{8.1} & \num{9.4 \pm 0.2} & \num{5.6} \\
    $\epsilon_y$ (nm) & \num{3.21} & \num{3.22 \pm 0.02} & \num{0.6} \\
    $\beta_y$ (m) & \num{4.15} & \num{4.16 \pm 0.03} & \num{0.4} \\
    $\alpha_y$  & \num{0.32} & \num{0.31 \pm 0.01} & \num{-0.6} \\
    $\eta_y$ (mm) & \num{-1.3} & \num{-1.5 \pm 0.1} & \num{-1.9} \\
    $\eta_{y'}$ (\num{e-3}) & \num{1.0} & \num{1.5 \pm 0.2} & \num{3.1} \\
    $\theta_{xy}$ (mrad) & \num{99.95} & \num{99.67 \pm 0.10} & \num{-2.9} \\
    $\sigma_p$ (\num{e-3}) & \num{0.96} & \num{0.96 \pm 0.02} & \num{-0.4} \\
    \bottomrule
    \end{tabular}
    \caption{Target and reconstructed beam parameters for the BTS simulation study. The target parameters were generated by applying 5\% random errors and a \qty{0.1}{rad} tilt to the design values. The reconstruction uncertainties are estimated from 5 different optimizations with slightly varied initial guesses. The last column reports the difference between the reconstructed and target values in units of the reconstruction uncertainty.
    }
    \label{tab:sim-recon-result}
\end{table}

\begin{figure}[htb!]
    \centering
    \includegraphics[width=\linewidth]{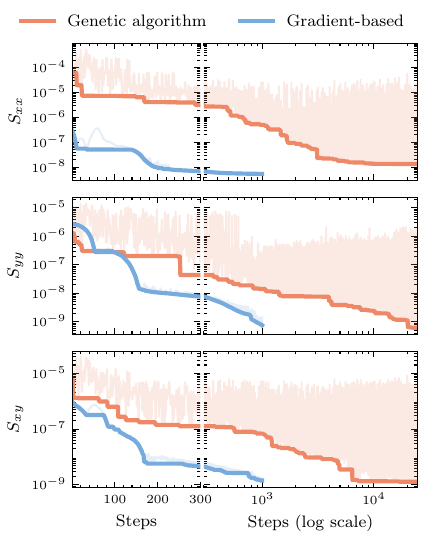}
    \caption{Optimization progress of using gradient descent compared to genetic algorithm. The $S_{xx}, S_{yy}, S_{xy}$ are the maximum errors of the beam transverse second moments across the 30 evaluated quadrupole settings respectively. The shaded lines denote the individual evaluations and the solid lines denote the best achieved errors so far.}
    \label{fig:gd_vs_ga_comparison}
\end{figure}

In addition, we ran genetic algorithm on the same simulated task, using the \gls{CNSGA}~\cite{lu_chaotic_2013} implementation in the Xopt package. The progress of reconstruction errors of individual moments are shown in \cref{fig:gd_vs_ga_comparison}. The $S_{xx}, S_{yy}, S_{xy}$ stand for the maximum of the absolute errors across the 30 evaluated quadrupole settings of the beam transverse second moments respectively. Both methods reach similar final error levels of around \num{e-9} to \num{e-8} in terms of the second moments. As expected, the gradient-based optimization converges faster than the genetic algorithm, requiring only around \num{1000} evaluations which is an order of magnitude fewer than the steps required for the genetic algorithm. 

\section{Robustness and Challenges}\label{sec:robustness}

The performance of the reconstruction method depends on both the quality of the measurements and the fidelity of the beamline model used in the optimization. In an idealized setting, the measured downstream beam moments are assumed to be fully determined by the initial beam sigma matrix and the known lattice model. 
In practice, however, several effects can introduce deviations between the simulated and measured beam profiles. We consider three error sources in the following, including uncertainty in the extracted beam moments, systematic errors in magnet strengths, and modeling errors associated with unaccounted beam offsets.

\subsection{Finite Screen Resolution}

\begin{figure}[htb!]
    \centering
    \includegraphics[width=0.85\linewidth]{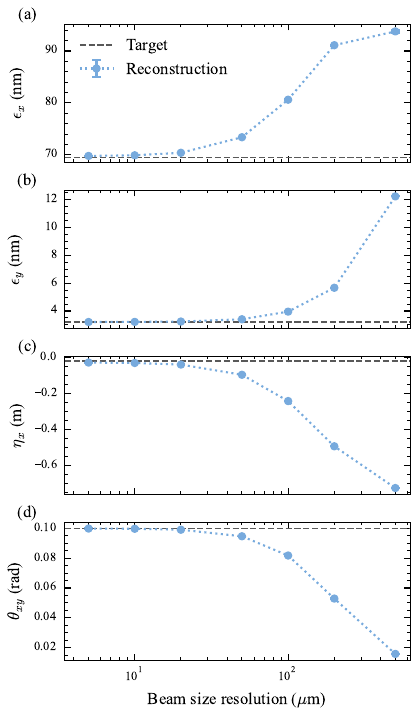}
    \caption{Reconstructed parameters compared to their target values depending on effective beam size resolution.}
    \label{fig:screen-resolution-sensitivity}
\end{figure}

First, the finite screen resolution imposes a lower bound on the beam profile measurement uncertainties. This effect can be simulated by adding a quadratic error term $\sigma_\text{res}^2$ to the measured second moments, where
$\sigma_\text{res} = \sigma_\text{pixel} /\sqrt{12}$.
\Cref{fig:screen-resolution-sensitivity} shows the reconstructed beam parameters with different effective screen resolutions.
The reconstruction errors of the individual beam parameters remain under a few percent when the resolution is below \qty{20}{\micro\meter}, corresponding to a pixel size of \qty{70}{\micro\meter}. As the screen resolution becomes comparable to the beam size, the inferred second moments become increasingly biased and significantly reduce the reconstruction accuracy.
In the case of the \gls{BTS} line, the pixel sizes are around \qty{20}{\micro\meter}, which is estimated to be sufficient for determining the upstream beam parameters.

\subsection{Systematic Magnet Strength Errors}

\begin{figure}[htb!]
    \centering
    \includegraphics[width=0.85\linewidth]{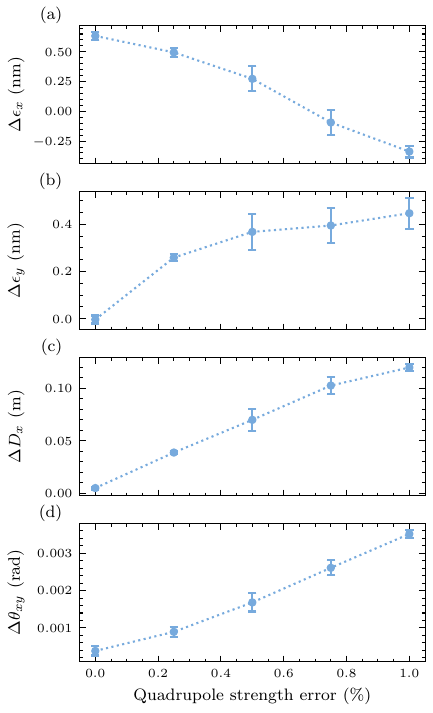}
    \caption{Errors in the reconstructed parameters depending on systematic error of quadrupole magnetic strengths.}
    \label{fig:strength-sensitivity}
\end{figure}

Another source of systematic error is the uncertainty in the magnet strengths. The quadrupole strength errors can lead to miscalibration of the beam transport model and therefore bias the reconstructed beam parameters. This effect is simulated by adding a percentage systematic error term $\sigma_\text{quad}$ to all the quadrupole strengths. The results are shown in \cref{fig:strength-sensitivity}. 

As the quadrupole strength errors increase, the reconstructed parameters deviate more from the target values. However, this effect impacts different beam parameters differently. 
For example in \cref{fig:strength-sensitivity}.~(a), the emittance is initially overestimated. As the assumed systematic strength error increases, the reconstructed emittance decreases, causing the error to first become smaller before increasing again once it is underestimated. This illustrates that the systematic error could partially compensate other uncertainty and biases, but should not be considered as a improvement of the model.
Overall, for parameters with larger values like the emittances, the relative errors remain under a few percent even with a 1\% strength error. For parameters with small values that are sensitive with respect to the beam size measurements, such as the dispersion $\eta_{x}$, the relative errors can exceed 100\% even with a small strength error. 

In principle, the quadrupole strength calibration factor could be included as an additional optimization. This would, however, make the inverse problem more brittle as the optimizer can trade-off errors in the simulation model against changes of the beam parameters. For this reason, this study uses fixed calibrated quadrupole strengths and does not include the calibration factor into the fit.
In practice, we expect that the quadrupole strengths and beam energy should be calibrated to better than 0.2\% for a sufficiently accurate reconstruction of the beam parameters. This can be achieved with dedicated magnet measurements and independent beam-based calibration methods.

\subsection{Magnet and beam offsets}

\begin{figure}[htb!]
    \centering
    \includegraphics[width=0.9\linewidth]{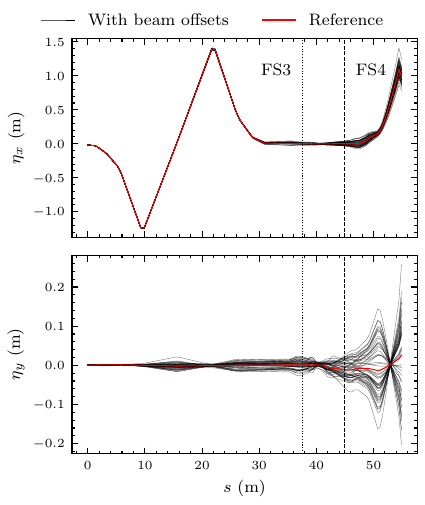}
    \caption{Dispersion functions along the BTS transport line for \gls{RMS} beam offsets of \qty{1}{mm}. The red line indicates the reference dispersion, with black lines showing the results from the randomly sampled beam offsets. The vertical dotted lines indicate the position of FS3 and FS4 screens used in this study.}
    \label{fig:dispersion_with_offsets}
\end{figure}

The magnet misalignment and beam offsets along the transport line introduce additional variable dispersions during the quadrupole scan measurement. During measurement, although the trajectory control is used to keep the beam roughly centered through the transport line, there will be remaining beam offsets at the quadrupole positions, as the \glspl{BPM} and correctors are positioned only at several locations along the line.
\Cref{fig:dispersion_with_offsets} shows the resulting dispersion functions $\eta_{x,y}$ with \qty{1}{mm} \gls{RMS} beam offsets at quadrupole locations, with the red line showing the reference dispersion. The locations of the FS3 and FS4 screens are marked with vertical lines. The effect at the FS4 screen is expected to be more prominent compared to the one at the FS3 due to the larger original dispersion values. Overall, the variable dispersions are still small at the screen locations compared to the later stages of the transport line.

\begin{figure}[htb!]
    \centering
    \includegraphics[width=0.9\linewidth]{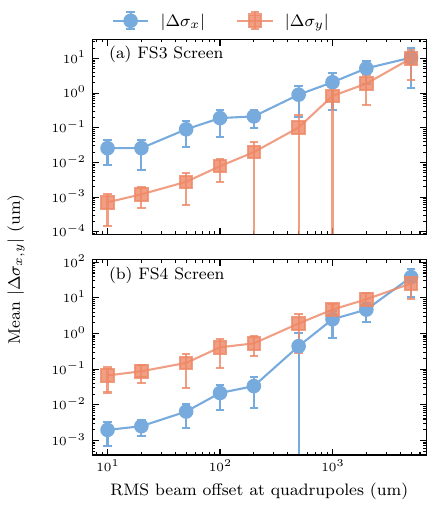}
    \caption{Effect of the magnet misalignment and beam offsets on the measured beam sizes at the FS3 and FS4 screens. The beam offsets at each magnet position are randomly sampled from \qty{10}{\micro\meter} to \qty{5}{mm}. The markers show the average estimated changes in the beam sizes in $x$ and $y$ for 30 different quadrupole settings, with the errorbar showing the uncertainty from 10 different sampled beam offsets.}
    \label{fig:beam-offset-beamsize-changes}
\end{figure}

The variable dispersion contributes to additional errors in the measured beam sizes. For an expected \qty{0.1}{\percent} energy spread, the effective changes in the measured beam sizes at the FS3 and FS4 screens are shown in \cref{fig:beam-offset-beamsize-changes}. The absolute changes in the beam sizes are averaged across 30 different quadrupole settings, with the error bars showing the variations from 10 different beam offset samples with \gls{RMS} values ranging from \qty{10}{\micro\meter} to \qty{5}{mm}. 
As expected, the changes in the beam sizes are more prominent at the FS4 screen due to the stronger dispersion. For an RMS beam offset of \qty{1}{mm}, the changes in the beam sizes remain under \qty{10}{\micro\meter} for both screens, which is comparable to the expected measurement uncertainty from the screen resolution. Although this effect should not significantly bias the reconstruction results, reaching sub-percent accuracy would require incorporating beam offsets and quadrupole misalignments into the reconstruction.

\section{Real-World Measurement at BTS Line}\label{sec:result_bts}

Several real-world measurements were conducted at the \gls{BTS} transport line to test the applicability of the method.
First, we performed a single-screen measurement using the FS3 screen. To mitigate the quadrupole strength errors due to hysteresis, we operate the quadrupoles always on the downramp of the major hysteresis loop and condition the magnets between steps if needed. The quadrupole scan settings are sorted to minimize the required measurement time.
During the measurement, the trajectory control was run intermittently to keep the beam approximately centered through the line. 
For each of the quadrupole settings, 10 subsequent beam images on the diagnostic screen were taken. Background was subtracted from the measured images, and the hot pixels were removed using a median filter. 
As the measured beam images contain saturated pixels, we extract the second moments by fitting a 2D Gaussian function to the thresholded pixels, between \qty{10}{\percent} and \qty{95}{\percent} of the maximum pixel intensity.

\begin{figure}[htb!]
    \centering
    \includegraphics[width=\linewidth]{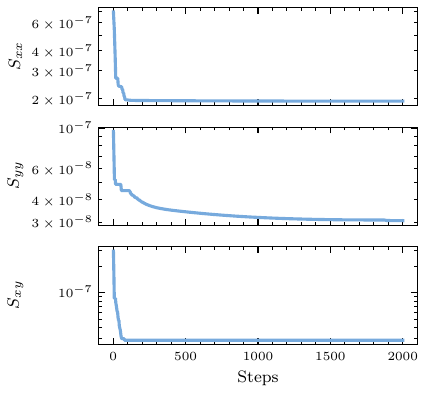}
    \caption{Optimization progress of the single-screen reconstruction. The $S_{xx}$, $S_{yy}$, and $S_{xy}$ are the maximum errors of the beam transverse second moments across the 30 evaluated quadrupole settings respectively.}
    \label{fig:single_screen_error}
\end{figure}

\begin{figure}[thb!]
    \centering
    \includegraphics[width=\linewidth]{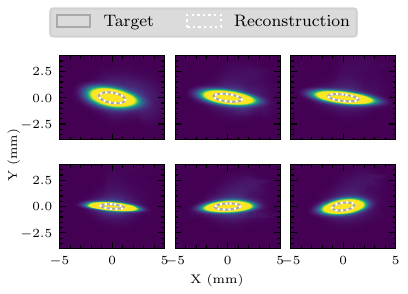}
    \caption{Example of the measured beam images on the FS3 screen, where each image represents one quadrupole setting. The target moments are marked with $1 \times \sigma$ gray ellipses and the reconstructed moments are shown as white dotted ellipses.}
    \label{fig:single_screen_images}
\end{figure}

In the reconstruction process, \num{1000} macroparticles were tracked, and the optimization was performed with Adam for \num{2000} steps. This corresponds to around 4 minutes of computation time on a laptop with Apple M-series chip. The optimization progress is shown in \cref{fig:single_screen_error}. The errors of the predicted second moments decrease rapidly in the first few hundred steps. The final error levels are around \num{e-8} to \num{e-7}, which is an order of magnitude higher than the simulated cases, due to the above-mentioned measurement uncertainties and modeling errors. 
Some measured beam images and the corresponding beam moments are shown in \cref{fig:single_screen_images}. The target moments are marked with $1 \times \sigma$ gray ellipses, and the reconstructed moments are marked with white dotted ellipses. The reconstructed moments agree well with the measured moments across the scan.

\subsection{Multi-Screen Measurement}\label{sec:multi_screen}

\begin{figure}[htb!]
    \centering
    \includegraphics[width=\linewidth]{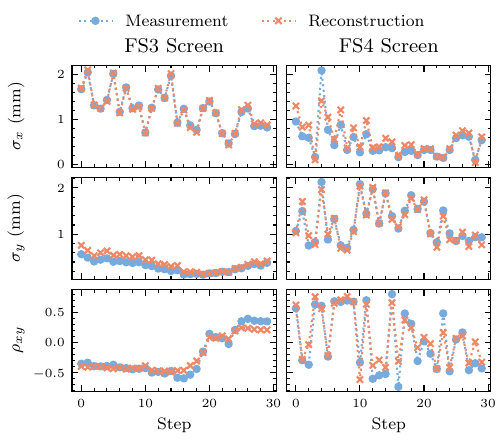}
    \caption{Transverse beam moments measured on the FS3 and FS4 diagnostic screens compared with the predicted moments from the reconstruction. The $\rho_{xy}$ is the dimensionless correlation coefficient. 
    }
    \label{fig:multi_screen_result}
\end{figure}

This measurement approach can be seamlessly generalized to incorporate measurements from different beamlines and diagnostic screens. At the \gls{BTS} transport line, the screen FS4 downstream of the emittance exchange section can be also used.
We conducted three independent measurements with different scan sets, each including 30 quadrupole settings. The maximum beam sizes are expected to be \qty{1.5}{mm} for Run1, \qty{2}{mm} for Run2, and \qty{2.5}{mm} for Run3.
In the reconstruction, the predicted moments on both screens are included in the loss function with equal weights. The optimization is performed with the same settings as the single-screen case.
\Cref{fig:multi_screen_result} shows the predicted beam moments compared with the measured moments for the Run2 results, where the $\rho_{xy}$ is the dimensionless correlation coefficient. In general, the predicted moments for FS3 agree better with the measured moments compared to the FS4 screen. This can be explained by a mismatched simulation model for the emittance exchange section.

The respective beam parameters from the three runs are listed in \cref{tab:multi_screen_meas_result}, along with the nominal parameter values for comparison.
The reconstruction results are broadly consistent across the three measurements: the horizontal emittance is about \qty{60}{\nano\meter}, and the vertical emittance is about \qty{1.2}{\nano\meter}. 
Notably, three runs also predict a horizontal dispersion of approximately \qty{-0.4}{\meter} and a transverse tilt angle of order \qty{-30}{mrad} to \qty{-50}{mrad}, which is not expected in the design settings. This may indicate an optics mismatch in the booster. Such a result could be further verified for example with standalone optics measurements in the booster ring~\cite{safranek1997experimental,sajaev2019lattice}. While this is not verified due to the limited beam time, it shows the potential of the method to identify optics errors in the upstream accelerator.

\begin{table}[thb!]
    \centering
    \caption{Results of the beam sigma matrix estimated from three multi-screen measurements compared with the nominal parameter values. Each scan used a different set of quadrupole settings. Run2 and Run3 were measured on the same day.
    }
    \label{tab:multi_screen_meas_result}
    \begin{tabular}{lcccc}
    \toprule
    Parameter & Run1 & Run2 & Run3 & Nominal \\
    \midrule
    $\epsilon_x$ (nm) & \num{60.69} & \num{61.92} & \num{59.90} & \num{70.21} \\
    $\beta_x$ (m) & \num{12.53} & \num{12.42} & \num{11.91} & \num{13.85} \\
    $\alpha_x$  & \num{2.29} & \num{2.31} & \num{2.23} & \num{2.61} \\
    $\eta_x$ (m) & \num{-0.39} & \num{-0.36} & \num{-0.40} & \num{-0.02} \\
    $\eta_{x'}$ (\num{e-3}) & \num{81.77} & \num{74.04} & \num{78.15} & \num{7.85} \\
    $\epsilon_y$ (nm) & \num{1.17} & \num{1.22} & \num{1.10} & \num{1.00} \\
    $\beta_y$ (m) & \num{3.31} & \num{2.91} & \num{3.20} & \num{3.15} \\
    $\alpha_y$  & \num{-0.95} & \num{-0.57} & \num{-0.83} & \num{-0.71} \\
    $\eta_y$ (mm) & \num{19.25} & \num{34.09} & \num{25.62} & \num{0.54} \\
    $\eta_{y'}$ (\num{e-3}) & \num{5.71} & \num{5.13} & \num{8.72} & \num{0.18} \\
    $\theta_{xy}$ (mrad) & \num{-31.37} & \num{-47.81} & \num{-35.38} & \num{-0.05} \\
    $\sigma_p$ (\%) & \num{0.12} & \num{0.13} & \num{0.13} & \num{0.10} \\
    \bottomrule
    \end{tabular}
\end{table}

As the inverse problem is nonlinear, the gradient-based optimization may converge to local minima. However, the consistency obtained from three independently selected scan sets suggests that the relevant minima are close in the physical beam-parameter space. This repeatability is an important indication that the multi-screen formulation improves the robustness of the reconstruction.

\begin{figure}[htb!]
    \centering
    \includegraphics[width=0.9\linewidth]{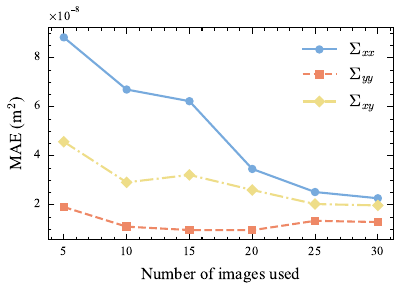}
    \caption{Effect of using a reduced number of measured beam images on the reconstructed beam moments. The error is calculated as the \gls{MAE} of the predicted beam moments across the 30 evaluated quadrupole settings compared to the measured moments.}
    \label{fig:error_vs_num_image_used}
\end{figure}

Incorporating multi-screen measurements could also reduce the number of required images, and therefore the beam time needed for a measurement. \Cref{fig:error_vs_num_image_used} shows the reconstruction result using only a subset of the measured images. As expected, the reconstruction error in the transverse beam moments increases as the number of images is reduced. This is consistent with the fact that, although the problem is formally overdetermined, the Jacobian can be ill-conditioned along certain directions in the parameter space, especially those with lower contribution to the beam sizes. The five-dimensional sigma matrix has 15 degrees of freedom, while each image provides three observables, so in principle only five images are required. In practice, however, additional measurements help better constrain correlations and reduce ambiguities in the reconstructed initial beam distribution, especially in the presence of measurement noise and systematic errors. 
For this study, using around 20 images appears to preserve most of the reconstruction accuracy. This should be regarded as an initial estimate, since the current result is based only on down-selecting from an existing dataset. If fewer images are to be acquired in practice, the quadrupole settings could instead be re-optimized to maximize the information content of the measurement set.
The computational speed of the reconstruction is comparable to the time required for magnet conditioning and image acquisition during the quadrupole scan. This opens the possibility of performing the reconstruction online during the measurement and using the intermediate result to adaptively select subsequent scan settings. In future implementations, we plan to develop an adaptive workflow incorporating optimal experimental design criteria to choose the next setting based on the current reconstructed beam matrix and its uncertainty, thereby improving measurement efficiency while preserving reconstruction accuracy.

\section{Comparison with GPSR}\label{sec:gpsr}

\begin{figure*}[htb!]
    \centering
    \includegraphics[width=0.85\linewidth]{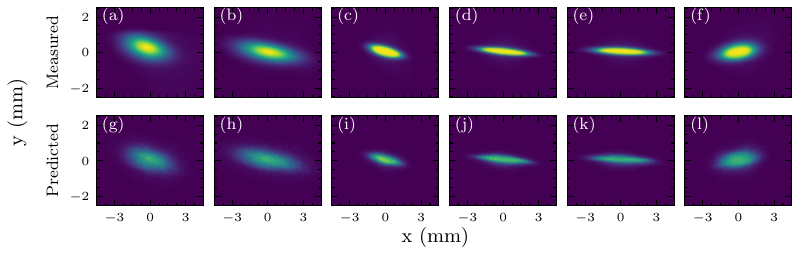}
    \caption{Comparison of measured transverse beam profiles (top row, a-f) and predicted images (bottom row, g-l) based on the reconstructed beam distribution from the \gls{GPSR} method. Each column corresponds to a different quadrupole scan setting. This data corresponds to Run1 in~\cref{tab:multi_screen_meas_result}.}
    \label{fig:gpsr_image_comparison}
\end{figure*}

The real-world measurements above show that quadrupole scans in the \gls{BTS} line can provide enough information to consistently reconstruct a five-dimensional beam sigma matrix using only second moments of the measured screen profiles. 
The same measurement concept also provides a natural connection to \gls{GPSR}, which uses differentiable tracking and a generative model like a fully-connected \gls{NN} to reconstruct the detailed phase-space information rather than only its covariance matrix~\cite{roussel2023phase,roussel2024efficient}.

The two approaches therefore address different levels of the same diagnostic problem. The sigma-matrix method used here is well suited for routine measurements when the beam can be sufficiently described by its second moments. In fact, the sigma-matrix reconstruction can be regarded as a special case of \gls{GPSR} where the generative model is constrained to linear transformation of a multivariate Gaussian distribution. This makes the sigma-matrix method more computationally efficient and less prone to overfitting when the beam distribution is close to Gaussian.
More generally, \Gls{GPSR} can represent non-Gaussian structure and higher-order correlations that are not contained in the sigma matrix, such as local density variations, but it generally requires greater computational effort. This makes \gls{GPSR} a useful extension when the measured profiles indicate complex phase-space structure or when downstream prediction requires more than a \gls{RMS} description. 

To compare the performance of the two methods, we applied \gls{GPSR} to the same quadrupole-scan dataset used for the sigma-matrix reconstruction. A 3-layer fully-connected \gls{NN} with 128 neurons per layer was used as the generative model for the incoming beam distribution. The predicted screen images from the \gls{GPSR} reconstruction are compared with the measured images in \cref{fig:gpsr_image_comparison}. The \gls{GPSR} results also show good agreement with the measured profiles and remain approximately Gaussian, which is to be expected for the booster extracted beam. However, it did not reach the same level of accuracy as the scalar reconstruction method based on the second moments. One important reason is the saturated screen images, which are currently not accounted for in the \gls{GPSR} model as it would require adaptive thresholding of the generated screen images and complicates the optimization process. In comparison, the sigma-matrix method was not affected as it uses the Gaussian distribution fitted only to thresholded pixel intensities. 
In addition, the optimization of the \gls{NN} generative model is more computationally intensive and more prone to local minima compared to the sigma-matrix reconstruction, which may contribute to the reduced accuracy.

\section{Conclusions}
We have proposed and developed a novel method for reconstructing the five-dimensional beam sigma matrix in a transport line from quadrupole-scan measurements using the differentiable accelerator simulation code Cheetah. The method directly optimizes the unknown beam sigma matrix against the transverse beam moments measured downstream, providing a computationally efficient and physics-consistent approach to beam characterization in dispersive transport lines.

The method was demonstrated experimentally at the \gls{APS} \gls{BTS} transport line, where the reconstructed sigma matrices showed good consistency across independent measurements. Simulation studies were used to evaluate practical factors that can affect the reconstruction accuracy, including diagnostic resolution, systematic quadrupole-strength errors, beam offsets, and unmodeled magnet offsets.

Overall, the results demonstrate the feasibility of differentiable simulation-based beam matrix reconstruction for practical transport-line characterization. The flexible formulation is expected to generalize to other transport lines and measurement configurations. In addition, the method can be extended to perform full 6D beam matrix reconstruction with available longitudinal beam diagnostics like \glspl{TCAV}.
The beam sigma matrix reconstruction framework also seamlessly integrates into \gls{GPSR} for resolving more complex phase-space structures, which is especially relevant for detailed beam characterization in linear accelerators. 
The proposed method provides a natural foundation for future online deployment in automated beam characterization workflows~\cite{roussel2026autonomous}, as well as further extensions toward uncertainty quantification, adaptive measurement selection, integrated system identification, and model-based accelerator tuning.

\section*{Code availability}

The code for the differentiable beam sigma-matrix reconstruction along with the processed data is available at \url{https://github.com/cr-xu/diff-5d-sigma-matrix}. Raw data can be made available upon request.

\begin{acknowledgments}
  The work is supported by the U.S. DOE Office of Science-Basic Energy Sciences, under Contract No. DE-AC02-06CH11357.
  We gratefully acknowledge the computing resources provided on Swing, a high-performance computing cluster operated by the Laboratory Computing Resource Center at Argonne National Laboratory.
  Generative AI tools were used to assist with language refinement and phrasing. All scientific content, analysis, and conclusions were reviewed and verified by the authors.
\end{acknowledgments}


\bibliography{bibliography} 

\appendix


\end{document}